\documentclass{article}
\usepackage{spconf,amsmath,graphicx}
\usepackage[pdftex]{color}
\usepackage{hyperref}
\usepackage{comment}
\usepackage{url}
\hypersetup{
    colorlinks = false,
}

\renewcommand{\Vec}[1]{\textrm{\boldmath $#1$}} % Vector

\newcommand{\revise}[1]{\textcolor{black}{#1}}
\title{Low-latency Incremental Text-to-Speech Synthesis with \\Distilled Context Prediction Network}
\name{Takaaki Saeki, Shinnosuke Takamichi, and Hiroshi Saruwatari}
\address{Graduate School of Information Science and Technology, The University of Tokyo, Japan.}

\begin{document}
%\ninept
%
\maketitle
\begin{abstract}
Incremental text-to-speech (TTS) synthesis generates utterances in small linguistic units for the sake of real-time and low-latency applications.
We previously proposed an incremental TTS method that leverages a large pre-trained language model to take unobserved future context into account without waiting for the subsequent segment.
Although this method achieves comparable speech quality to that of a method that waits for the future context, it entails a huge amount of processing for sampling from the language model at each time step. 
\revise{In this paper, we propose an incremental TTS method that directly predicts the unobserved future context with a lightweight model, instead of sampling words from the large-scale language model.}
We perform knowledge distillation from a GPT2-based context prediction network into a simple recurrent model by minimizing a teacher-student loss defined between the context embedding vectors of those models.
Experimental results show that the proposed method requires about ten times less inference time to achieve comparable synthetic speech quality to that of our previous method, \revise{and it can perform incremental synthesis much faster than the average speaking speed of human English speakers, demonstrating the availability of our method to real-time applications.}
\end{abstract}

\vspace{-1mm}
\section{Introduction}
\vspace{-1mm}
\revise{
Text-to-speech (TTS) synthesis, which artificially synthesizes speech from text, can now generate extremely human-like and high-quality speech with the recent development of deep learning~\cite{shen17tacotron2,ren2021fastspeech,weiss21wave,donahue2021endtoend}.
Most of these TTS methods are ``sentence-level TTS'' methods, which take an entire input sentence into account with an encoder-decoder model to generate the acoustic feature of the output speech.
However, some applications, such as simultaneous speech-to-speech translation~\cite{bangalore-etal-2012-real,sudoh2020simultaneous,zheng-etal-2020-fluent} or real-time dialog speech generation~\cite{bertero-etal-2016-real}, need streaming synthesis, namely ``incremental TTS'', that processes small linguistic units one by one rather than sentence-level TTS.
Although incremental TTS can immediately process the input text, its synthetic speech quality tends to be lower than sentence-level TTS.
The biggest challenge of incremental TTS is to naturally synthesize the current segment from only the observed context information without enough subsequent context information.
Incremental TTS that takes only the past observed information into account~\cite{Yanagita19itts} can efficiently synthesize the current speech segment without any waiting time, but it tends to output significantly unnatural and discontinuous speech compared with sentence-level TTS.
}
The lookahead-$k$ policy~\cite{Ma20prefix}, which waits a word sequence of length $k$ before generating the current segment, successfully copes with the quality degradation, but it suffers from latency entailed by waiting for the subsequent text segments at every time step.

We previously proposed an incremental TTS method that leverages a large pre-trained language model to take future context into account by using only the observed segment~\cite{saeki20itts_lm} (hereafter, ``the conventional method'').
This method uses subsequent text segments generated with GPT2~\cite{radford2019language} (hereafter, ``pseudo lookahead'') instead of waiting for the unobserved segment as in the lookahead-$k$ policy. 
Although this method can improve the naturalness of incremental TTS without waiting for the observation of the subsequent segment, the inference time of GPT2 is critical for low-latency incremental TTS especially when computational resources are limited.
Therefore, for practical applications, it is necessary to predict the unobserved future context from the observed information by using only a lightweight model, which is generally a challenging task (see Section~\ref{sec:discussion}). 

In this paper, we propose a low-latency incremental TTS method that predicts the unobserved future context with a lightweight model.
This method directly predicts the unobserved future context needed for incremental TTS based on the linguistic knowledge of the large pre-trained language model without heavy computation of the inference process.
\revise{On the basis of the recent success of task-specific knowledge distillation of a large-scale language model in the field of natural language processing~\cite{chia2019TransformerTC,tang2019distilling,chen2020adabert}, the proposed method leverages the teacher-student training to obtain a lightweight context prediction network from the GPT2-based model.}
The teacher model used in this paper is based on the incremental TTS model of our previous work~\cite{saeki20itts_lm}, with some improvements.
The student model is an incremental TTS model which has a simple recurrent network consisting of a single-layer bi-directional long-short-term memory (BLSTM) as the distilled context prediction network.
\revise{
Experimental results show that the proposed method 1) achieves the equivalent inference speed and much higher synthetic speech quality compared to the method without the subsequent information, 2) attains ten times faster inference to comparable synthetic speech quality to that of the conventional method, and 3) performs incremental synthesis much faster than the average speaking speed of human English speakers, demonstrating the availability of our method to real-time applications.
}

\vspace{-1mm}
\section{Related work}
\vspace{-1mm}
A number of TTS methods based on neural sequence-to-sequence models have been proposed~\cite{wang17tacotron,shen17tacotron2,Nihan19transformer,ren19fastspeech,ren2021fastspeech}.
Our method is based on Tacotron2~\cite{shen17tacotron2}, with some modifications to formulate incremental TTS.
Many recent streaming, incremental TTS studies also leverage neural sequence-to-sequence models~\cite{Yanagita19itts,Ma20prefix,Stephenson2020,Mohan2020IncrementalTT,Ellinas2020,Stephenson2021AlternateEI,he2021multi}. 
Yanagita et al. proposed an incremental TTS method using Tacotron~\cite{Yanagita19itts}.
This method was the first neural sequence-to-sequence incremental TTS, in which each input text segment is processed independently of the pre- and post-contexts, which limits the naturalness of output speech.
Ma et al. proposed prefix-to-prefix decoding, which waits for the next $k$~words at every time step of incremental generation~\cite{Ma20prefix}.
Although this method takes contexts into account, as does our method, the training process uses the entire utterance and the inference process is not segmental, in contrast to our method.
We previously proposed an incremental TTS method with a mechanism for generating unobserved context by using GPT2~\cite{saeki20itts_lm}, which we call ``the conventional method'' in this paper.
In contrast to the conventional method, the method proposed here does not use GPT2 during inference and uses a simple distilled context prediction network to obtain the context embedding, which significantly reduces the processing time.
\revise{To achieve highly accurate context prediction with a lightweight model, we leverage a knowledge distillation approach~\cite{Ba14DoDN,hinton14distil} where a smaller student model learns to mimic a larger teacher model.
There have been several studies on task-specific knowledge distillation of large-scale language models in the field of natural language processing~\cite{chia2019TransformerTC,tang2019distilling,chen2020adabert}.
In particular, a study on distilling BERT~\cite{devlin2019bert} into a single-layer BLSTM~\cite{tang2019distilling} showed that the student model achieved comparable performance to the teacher model on several benchmark tasks.}
Our work focuses on the task of context estimation in TTS using a similar recurrent architecture for the student model.

\begin{figure}[t]
  \centering
  \includegraphics[width=1.0\linewidth, clip]{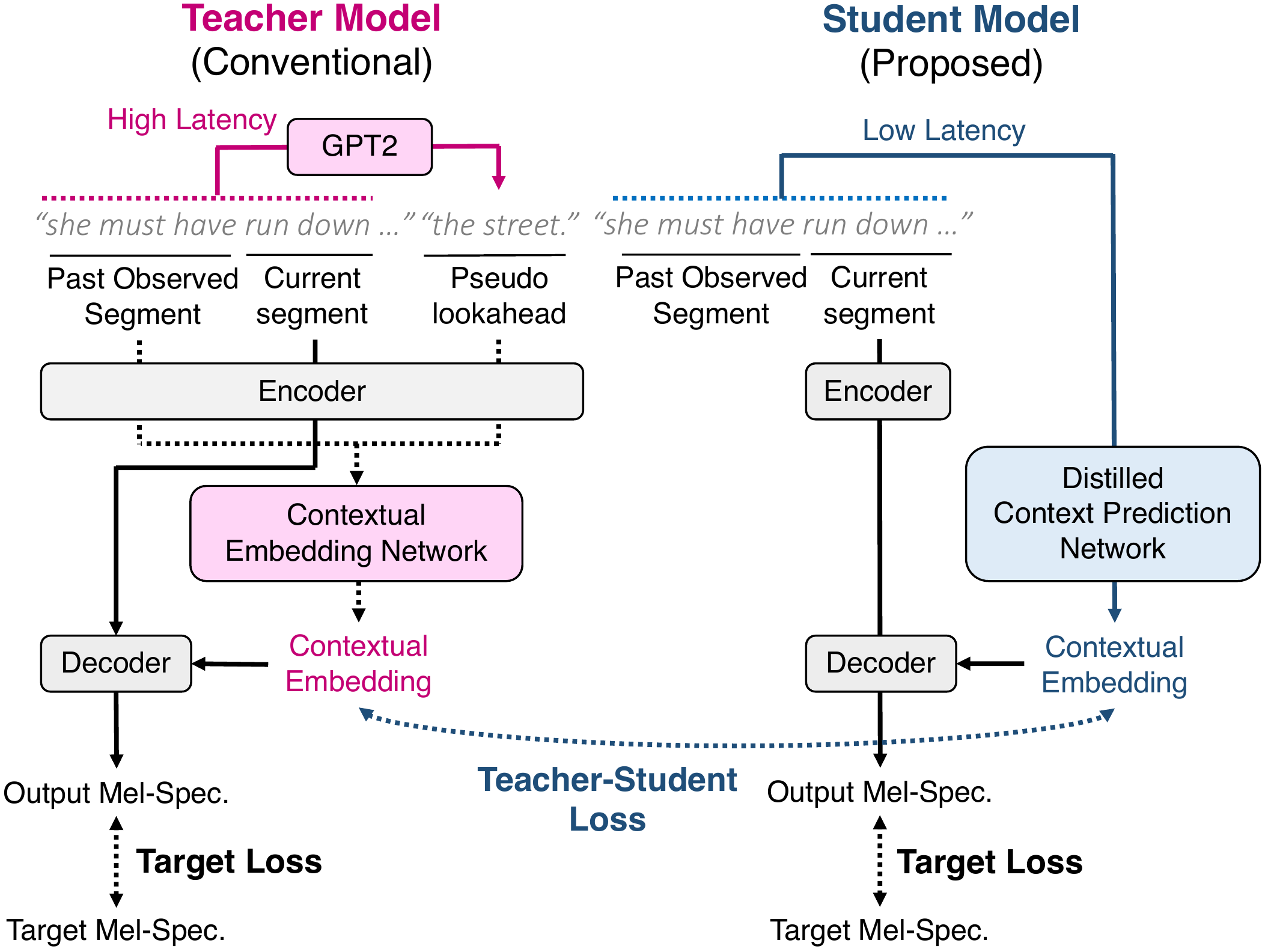}
  \caption{Proposed incremental TTS method with distilled context prediction network. Student model can be obtained by distilling GPT2-based context prediction model into lightweight recurrent network with teacher-student training.}
  \label{fig:concept}
\end{figure}

\vspace{-1mm}
\section{Conventional method}\label{sec:conventional}
\vspace{-1mm}
This section describes the conventional incremental TTS method~\cite{saeki20itts_lm} that uses GPT2~\cite{radford2019language} and Tacotron2~\cite{shen17tacotron2}'s encoder-decoder architecture, as shown on the left side of Fig.~\ref{fig:concept}.
Let $\Vec{w}_{n}$ be the $n$-th word of the input sentence, where each input segment consists of $N$ words.
$\Vec{w}_{1:Nt} = \Vec{w}_{1}, \cdots, \Vec{w}_{n}, \cdots, \Vec{w}_{Nt}$ represents ``the observed segment'' and the last $N$-word sequence $\Vec{w}_{N(t-1)+1:Nt} = \Vec{w}_{N(t-1)+1}, \cdots, \Vec{w}_{Nt}$ is ``the current segment'' in the time step $t$.
GPT2 assumes that the probability distribution of an $M$-word sequence $\Vec{w}_{1:M}$ can be decomposed into the product of conditional probabilities, as
\begin{align}
    p(\Vec{w}_{1:M}) = {\displaystyle \prod_{m=1}^{M}}p(\Vec{w}_{m}|\Vec{w}_{1:m-1}).
\end{align}
We can obtain a future $L$-word sequence $\hat{\Vec{w}}_{Nt+1:Nt+L} = \hat{\Vec{w}}_{Nt+1}, \cdots, \hat{\Vec{w}}_{Nt+L}$ by sampling from the probability distribution $p(\Vec{w}_{Nt+1:Nt+L} | \Vec{w}_{1:Nt})$, where $\hat{\Vec{w}}_{Nt+1:Nt+L}$ becomes the pseudo lookahead for the future contextual information of incremental TTS.
Then, we estimate the contextual embedding $\Vec{e}_{\mathrm{pseudo}}$ from a hidden state of the past observed segment $\Vec{h}_{1:N(t-1)}$ and that of the pseudo lookahead $\hat{\Vec{h}}_{Nt+1:Nt+L}$, as
\begin{align}
    \Vec{e}_{\mathrm{pseudo}} &= E(\Vec{h}_{1:N(t-1)}, \hat{\Vec{h}}_{Nt+1:Nt+L}) \\
    \hat{\Vec{h}}_{Nt+1:Nt+L} &= encoder(\hat{\Vec{w}}_{Nt+1:Nt+L}) \\
    \Vec{h}_{1:N(t-1)} &= encoder(\Vec{w}_{1:N(t-1)}),
\end{align}
where $E(\cdot)$ denotes a contextual embedding network, which consists of a style token layer~\cite{wang2018style}.
Then, we obtain the output mel-spectrogram $\Vec{y}_{N(t-1)+1:Nt}$ by conditioning the decoder on $\Vec{e}_{\mathrm{pseudo}}$, as
\begin{align}\label{eq:current_mel}
    \Vec{y}_{N(t-1)+1:Nt} &= decoder(\Vec{h}_{N(t-1)+1:Nt}, \Vec{e}_{\mathrm{pseudo}}) \\
    \Vec{h}_{N(t-1)+1:Nt} &= encoder(\Vec{w}_{N(t-1)+1:Nt}).
\end{align}
After that, we jointly train the encoder, decoder, and $E(\cdot)$ to minimize an objective function $\mathcal{L}_{\mathrm{target}}$, which can be calculated from the ground-truth mel-spectrogram and the output mel-spectrogram $\Vec{y}_{N(t-1)+1:Nt}$. $\mathcal{L}_{\mathrm{target}}$ consists of the mean squared error (MSE) loss and the stop flag prediction loss, in the same manner as Tacotron2.

The above method enhances the quality of synthetic speech by using pseudo lookahead without waiting for the observation of the future context.
However, the latency in synthesis significantly increases as the time step $t$ increases because of the high computational cost of GPT2's inference.

\vspace{-1mm}
\section{Proposed method}
\vspace{-1mm}

\revise{
This section describes the proposed incremental TTS method.
Section~\ref{sec:method_distil} presents the teacher-student training to obtain the lightweight distilled context prediction network.
Section~\ref{sec:method_imp} describes the implementation details of the teacher and student models.
In Section~\ref{sec:discussion}, we discuss the proposed method in detail based on the results of preliminary experiments.
}

\subsection{Knowledge distillation of context prediction network}\label{sec:method_distil}
Fig.~\ref{fig:concept} illustrates the proposed knowledge distillation method.
The context prediction network with GPT2 is distilled into a simple recurrent network (hereafter, ``the distilled context prediction network'') on the basis of teacher-student training.
The teacher model estimates the contextual embedding $\Vec{e}_{\mathrm{pseudo}}^{(T)}$ by using GPT2 and the contextual embedding network $E(\cdot)$, as shown in equation~(2)--(4).
In the student model, the distilled context prediction network directly predicts the contextual embedding from the observed sentence.
We use a pre-trained word embedding to obtain the distributed representation of the input text, as in the previous study~\cite{tang2019distilling}.
We apply a pre-trained FastText~\cite{fasttext17} to each word of the observed segment $\Vec{w}_{1:Nt}$ and obtain a word embedding sequence $\Vec{v}_{1:Nt}$.
When we define the distilled context prediction network of the student model as $G^{(S)}(\cdot)$, the contextual embedding $\Vec{e}_{\mathrm{pseudo}}^{(S)}$ is estimated as
\begin{align}
    \Vec{e}_{\mathrm{pseudo}}^{(S)} = G^{(S)}(\Vec{v}_{1:Nt}).
\end{align}
The objective of the teacher-student training is the MSE of the two contextual embedding vectors, which is formulated as
\begin{align}
    \mathcal{L}_{\mathrm{distil}} = ||\Vec{e}_{\mathrm{pseudo}}^{(S)} - \Vec{e}_{\mathrm{pseudo}}^{(T)} ||_{2}^{2}.
\end{align}
In our preliminary experiment, we also tried the cosine similarity loss for $\mathcal{L}_{\mathrm{distil}}$, and the MSE showed slightly better results.
To avoid degrading the naturalness of output speech when distilling the context prediction network, we also used the target loss $\mathcal{L}_{\mathrm{target}}$, which can be calculated from the ground-truth mel-spectrogram and the output mel-spectrogram of the student model, as described in Section~\ref{sec:conventional}.
The whole training objective is formulated as
\begin{align}\label{eq:whole_objective}
   \mathcal{L} =  (1-\lambda) \cdot \mathcal{L}_{\mathrm{target}} + \lambda \cdot \mathcal{L}_{\mathrm{distil}},
\end{align}
where $\lambda$ denotes the weight of the teacher-student loss.
Since the training process aims at obtaining a network that predicts the unobserved context from the observed segment, the student model inherits and fixes the weights of the encoder and decoder from the teacher model and only updates the weights of the distilled context prediction network $G^{(S)}$. 

\subsection{Implementation details}\label{sec:method_imp}
The architecture of the TTS model that estimates the mel-spectrogram from text is the same for the teacher and student models, and it is identical to the architecture of the conventional method.
After the encoder-decoder model outputs the mel-spectrogram from the current segment as in equation~\ref{eq:current_mel}, the WaveGlow Vocoder~\cite{prenger2018waveglow} synthesizes the waveform segment and the output speech is obtained by combining it with the segments synthesized so far.
As in the study~\cite{Ma20prefix}, the last $\delta$ frame of the mel-spectrogram of each segment is copied and padded to the end of the mel-spectrogram before synthesizing the waveform to alleviate the discontinuity of the concatenation.
As a preliminary experiment, we varied $\delta$ from zero to 20 and objectively compared the synthetic speech qualities as in Section~\ref{sec:evaluation_obj}.
On the basis of this evaluation, we decided to use $\delta = 1$ as it gave the lowest character error rate (CER).
\revise{Moreover, we set the number of words in each segment $N$ to two, in the same manner as our prior work~\cite{saeki20itts_lm}, since our preliminary studies found that all the methods described in Section~\ref{sec:evaluation} often generate unintelligible speech with $N=1$, as reported in Yanagita et al.'s work~\cite{Yanagita19itts}.}

\begin{figure}[t]
  \centering
  \includegraphics[width=1.0\linewidth, clip]{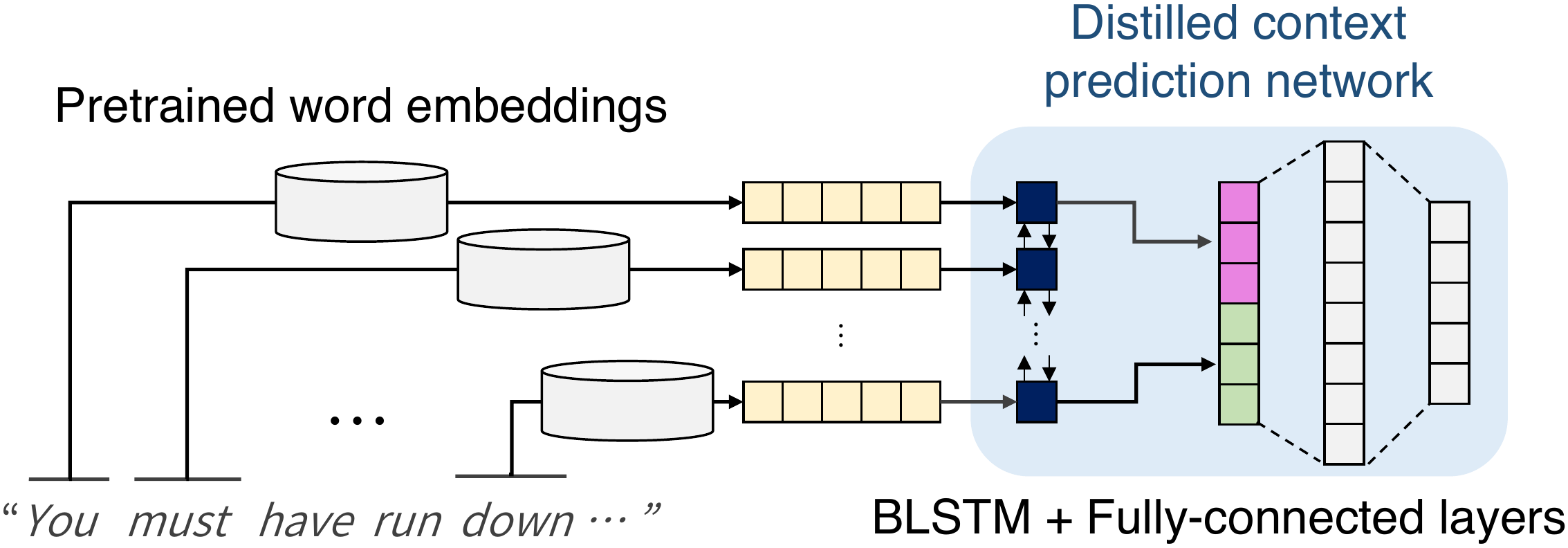}
  \caption{Model architecture of distilled context prediction network. Model consists of BLSTM layer and fully-connected layers, and it predicts contextual embedding from word embedding sequence of observed text.}
  \label{fig:model}
\end{figure}

The teacher model's contextual embedding network $E(\cdot)$ has the same architecture as the conventional method.
When sampling with GPT2, we set the maximum sampling length to $L=5$ and stop sampling when the end of a sentence is sampled, instead of always sampling at $L=5$ as in our prior work~\cite{saeki20itts_lm}.
This operation generates a context embedding with the pseudo lookahead that is closer to using the ground-truth lookahead.

The student model consists of a distilled context prediction network that predicts $\Vec{e}_{\mathrm{pseudo}}^{(S)}$ from the observed segment.
We use three sizes of distilled context prediction network, i.e., ``small'', ``medium'', and ``large''.
Fig.~\ref{fig:model} illustrates the model architecture of the distilled context prediction network.
It consists of a single-layer BLSTM without attention, similar to the student model used in the previous work~\cite{tang2019distilling}.
The number of embedding dimensions of the word embedding sequence $\Vec{v}_{1:Nt}$ is 300, and the number of dimensions of the hidden state of the BLSTM is set to 100, 300, and 500 in the small, medium, and large models, respectively.
The initial and the last hidden states of the BLSTM layer are combined and passed to a fully-connected layer for mapping to 200, 600, and 1000--dimensional vectors in the small, medium, and large models, respectively.
After applying a ReLU activation function, it is passed through another fully-connected layer that maps it to a 256--dimensional vector $\Vec{e}_{\mathrm{pseudo}}^{(S)}$, which has the same number of dimensions as the teacher model's contextual embedding $\Vec{e}_{\mathrm{pseudo}}^{(T)}$.

\begin{figure}[t]
  \centering
  \includegraphics[width=0.95\linewidth, clip]{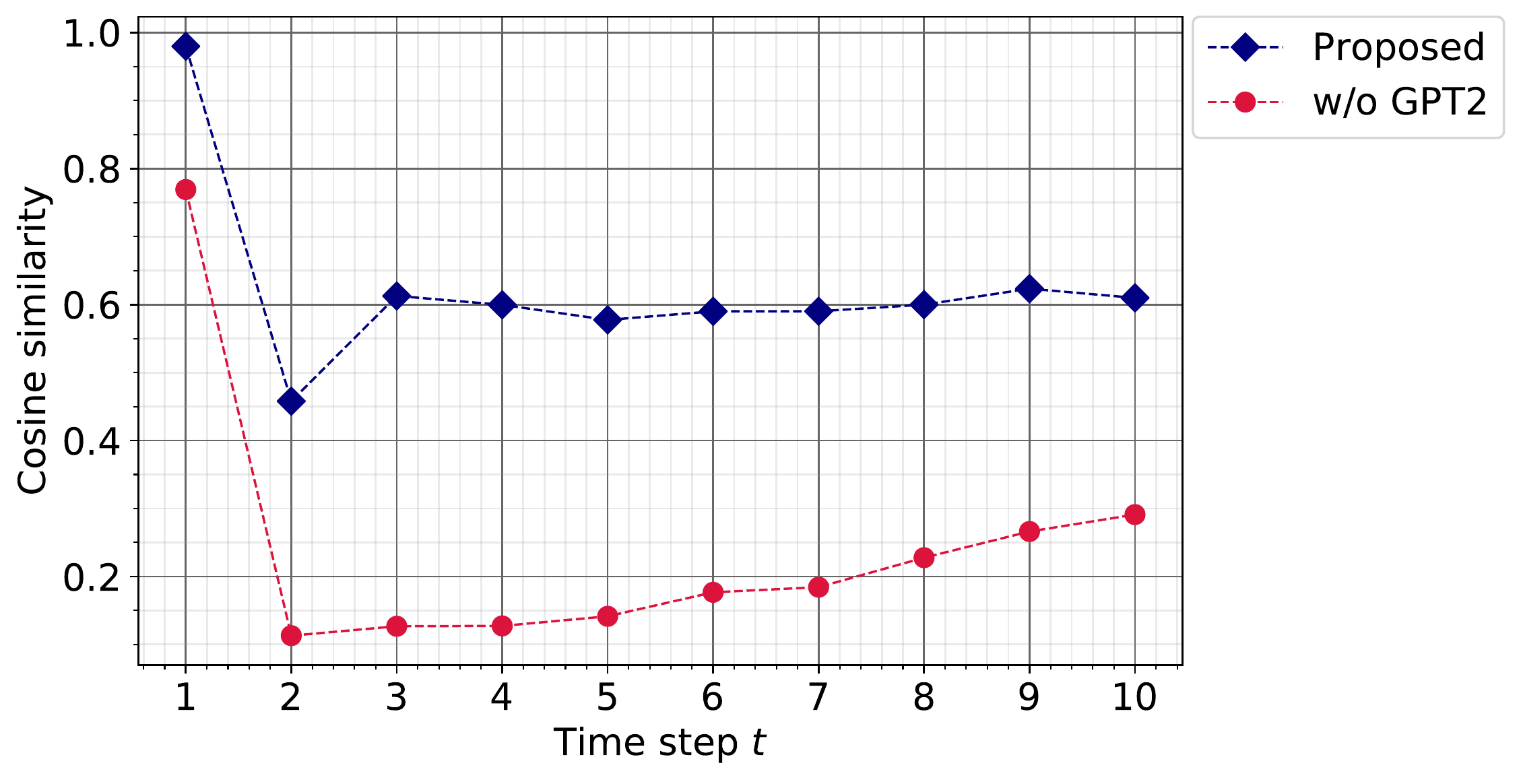}
  \caption{Average cosine similarity plots using medium distilled context prediction network with and without GPT2. Note that results for small and large distilled context prediction networks are almost identical to those for medium in this graph.}
  \label{fig:discuss}
\end{figure}

\subsection{Discussion}\label{sec:discussion}

First, let us investigate the necessity of GPT2 for training a student model with knowledge distillation.
The proposed method trains the student model to obtain the context embedding $\Vec{e}_{\mathrm{pseudo}}^{(T)}$ that the teacher model estimates using the pseudo lookahead generated with GPT2. 
On the other hand, there is an alternative method that does not use GPT2, which trains the student model to obtain the context embedding by using the teacher model and the ground-truth lookahead of the training data, instead of the pseudo lookahead. 
We calculated the cosine similarities between the embedding vectors of the teacher and student models at time step $t$ and compare them with the proposed method (``Proposed'') and the method without GPT2 (``w/o GPT2'').
Note that we used the objective function in eq.~\ref{eq:whole_objective} with $\lambda=1$ to evaluate only the accuracy of the context estimation.
Fig.~\ref{fig:discuss} shows the results.
``Proposed'' indicates the average cosine similarity of $\Vec{e}_{\mathrm{pseudo}}^{(S)}$ and $\Vec{e}_{\mathrm{pseudo}}^{(T)}$ at each time step $t$.
``w/o GPT2'' is the average cosine similarity of $\Vec{e}_{\mathrm{truth}}^{(S)}$ and $\Vec{e}_{\mathrm{truth}}^{(T)}$ for each time step $t$.
At high cosine similarity values, the student model can predict the contextual embedding of the teacher model with high accuracy.
The high cosine similarity at time step $t=0$ implies that context estimation is easier at the beginning of a sentence; our previous study found the same trend~\cite{saeki20itts_lm}.
As a result, we can conclude that ``Proposed'' significantly outperforms ``w/o GPT2'' for all time steps and model sizes.
This indicates that the student model obtained with the ground-truth sentences cannot be generalized to unknown test data and cannot predict the contextual embedding of the teacher model so well.
On the other hand, the student model distilled using GPT2 can predict the contextual embedding with high accuracy for the test data.

Now let us discuss the weight $\lambda$ in Eq.~\ref{eq:whole_objective}.
In the case of $\lambda=0$, the recurrent network simply estimates the contextual embedding for the TTS model instead of using GPT2's knowledge because the model uses only the target loss.
The case of $\lambda=1$ uses only the teacher-student loss $\mathcal{L}_{\mathrm{distil}}$ without the target loss $\mathcal{L}_{\mathrm{target}}$.
\revise{In our preliminary experiment, we found that $\lambda=0$ and $\lambda=0.5$ cases significantly deteriorated the stop flag prediction accuracy and often led to collapsed output speech.
Furthermore, since $\mathcal{L}_{\mathrm{distil}}$ is much smaller than $\mathcal{L}_{\mathrm{target}}$, $\lambda$ needs to be set to a value close to 1.0 in order to equalize the scale of these loss values.
Therefore, in the experimental evaluation of Section~\ref{sec:evaluation}, we chose $\lambda=0.95$ as the case of jointly using the teacher-student loss and the target loss, and we compared it with the case of $\lambda=1$ case.}

\vspace{-1mm}
\section{Experimental evaluations}\label{sec:evaluation}
\vspace{-1mm}
We conducted experimental evaluations by comparing the following methods: (1)~\textbf{\textit{Ground-truth}}, ground-truth audio clips included in the test data; (2)~\textbf{\textit{Full-sentence}}, sentence-level Tacotron2 model~\cite{shen17tacotron2}; (3)~\textbf{\textit{Teacher}}, the teacher model described in Section~\ref{sec:conventional}; (4)~\textbf{\textit{Student}}, the student model obtained with the proposed method; (5)~\textbf{\textit{Unicontext}}, the incremental TTS method used only the past observed segment with the contextual embedding network $E(\cdot)$; (6)~\textbf{\textit{Independent}}, the incremental TTS model that synthesizes the current speech segment independently~\cite{Yanagita19itts}.
\revise{As described in Section~\ref{sec:method_imp} and Section~\ref{sec:discussion}, we investigated three model sizes and $\lambda=\{0.95, 1.0\}$ for \textit{Student}.}
Note that we did not include \textit{Independent} in the evaluation of synthetic speech quality because its naturalness was significantly lower than those of other methods, as reported in our prior work~\cite{saeki20itts_lm}.
Audio samples with the above methods are available online\footnote{\scriptsize \url{https://takaaki-saeki.github.io/itts_distil_demo}}.

\subsection{Experimental conditions}
The dataset used for the evaluation was LJSpeech~\cite{ljspeech17}, which consists of 13,100~utterances (about 24 hours) by a single female English speaker.
We randomly selected 100 for the validation set and 500 utterances for the test set, and used the rest as the training set.
The sampling frequency was set to 22.05~kHz.
When extracting the mel-spectrogram from each audio clip, we used a 1024-sample frame size, 256-sample hop size, a Hann window function, and an 80~channel mel-filterbank. 

As pre-processing for training, we applied a sliding text window with a window length of three and a hop size of one, as in our previous study~\cite{saeki20itts_lm}.
During inference, the number of words in the input segment $\Vec{w}_{N(t-1)+1:Nt}$ was set to two.
The Kaldi-based toolkit~\cite{gentle17} was used to extract the ground-truth speech segments from the training data by using forced alignment.
For the evaluation, we used the published pre-trained GPT2\footnote{\scriptsize \url{https://github.com/graykode/gpt-2-Pytorch}} and WaveGlow\footnote{\scriptsize \url{https://github.com/NVIDIA/waveglow}}.
When training the teacher model, we performed GPT2-guided fine-tuning for 4000~iterations after consistently training the TTS model and the context embedding network for 76000~iterations in the same manner as our previous study~\cite{saeki20itts_lm}.
We performed the proposed knowledge distillation by setting the learning rate and batch size to $10^{-3}$ and 64, respectively.
We trained the distilled context prediction network for 20000~iterations by using the Adam~\cite{kingma14adam} optimizer with $\beta_{1} = 0.9$, $\beta_{2}=0.999$, and $\epsilon = 10^{-6}$.
We used Intel(R) Core(TM) i7-6850K @ 3.60~GHz CPUs and an NVIDIA GeForce 1080Ti GPU for measuring of the inference time.

\subsection{Objective evaluation of synthetic speech quality}\label{sec:evaluation_obj}
As an objective evaluation of naturalness, we conducted an evaluation using an automatic speech recognition model to calculate the CER and the word error rate (WER), defined as the character- and word-level Levenshtein distance~\cite{Levenshtein_SPD66}, by using the entire 500~utterances of the test data.
We used a joint-CTC Transformer-based model~\cite{kim2018jointctc} trained on Librispeech~\cite{librispeech}, which is included in ESPnet~\cite{espnet}.
Table~\ref{tab:evaluation_obj} lists the results.

\revise{
First, \textit{Unicontext}, which does not use any subsequent information, showed much higher CER and WER values than sentence-level TTS due to its unstable generation and discontinuous output speech.
} 
On the other hand, all of the \textit{Student} cases had significantly better scores than that of \textit{Unicontext}, demonstrating that the distilled context prediction network improves the quality of the synthetic speech more than when using only the observed information.
Furthermore, comparing the $\lambda=1$ and $\lambda=0.95$ cases of \textit{Student}, we can see that the CER for $\lambda=1$ is about 4~\% lower than that for $\lambda=0.95$, indicating that training with only the teacher-student loss gives better results.
\revise{This result is consistent with the previous study on the task-specific knowledge distillation of BERT for NLP tasks~\cite{tang2019distilling}, which reported that using only teacher-student loss achieved higher scores.
One possible reason is that, in the case of $\lambda=0.95$, the target loss causes the over-fitting of the distilled context prediction network to the target mel-spectrogram, instead of inheriting the context estimation ability of Teacher model.}
Moreover, looking at the results for each model size in \textit{Student} with $\lambda=1$, we can see that the scores are almost constant regardless of the model size.
Although the CERs for \text{Student} methods were lower than those of \textit{Teacher}, ``\textit{Student} (medium, $\lambda=1$)'' had the same WER score
, suggesting that the knowledge distillation learning can improve the score to the same level.

\subsection{Evaluation of inference time}\label{sec:evaluation_time}
\begin{figure}[tb]
  \centering
  \includegraphics[width=0.95\linewidth, clip]{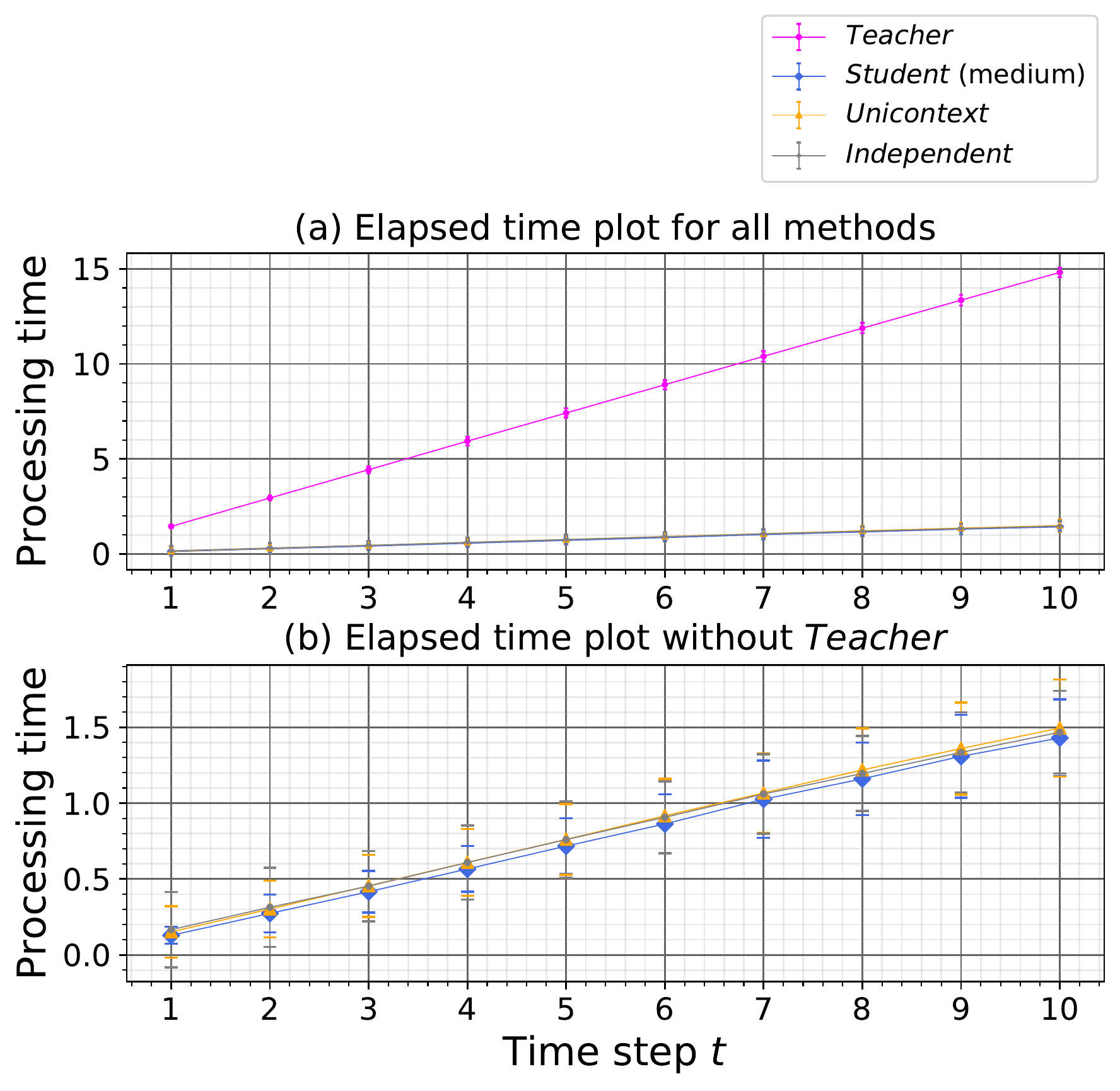}
  \caption{Cumulative processing time for time step $t$. (a) compares all incremental TTS methods, and (b) plots methods other than {\it Teacher}.}
  \label{fig:processing_time}
\end{figure}

To evaluate the processing time, we measured the cumulative processing time at each time step.
Fig.~\ref{fig:processing_time} shows the cumulative processing time for a time step $t$.

In Fig.~\ref{fig:processing_time}(a), we can see that \textit{Teacher} requires a significantly large amount of processing time for sentence generation at each time step, and it needs about 1.5~second to generate each segment.
On the other hand, \textit{Student}, which is obtained with the proposed knowledge distillation method, can efficiently generate each segment in a shorter processing time.
Fig.~\ref{fig:processing_time}(b) shows that the processing time of \textit{Student} is comparable to that of \textit{Independent} and \textit{Unicontext}, which do not use the future context.
The average amount of processing of \textit{Student} is even slightly smaller than that of \textit{Independent}, because \textit{Independent} sometimes fails to predict the stop flag and requires a large amount of time to finish the decoding process.

Previous studies on the reading rate of English speech have indicated that 100--120 words per minute (wpm) is optimal for an input utterance of simultaneous interpretation~\cite{Seleskovitch1978} and that the average speech rate for reading aloud is about 180~wpm~\cite{Brysbaert19}.
Therefore, the conventional method, which can synthesize output speech at around 80~wpm, is not sufficient for simultaneous speech-to-speech translation.
On the other hand, the proposed method has a generation speed of 800~wpm, which is much faster than the average English reading speed.

\begin{table}[tb]
\centering
\caption{Evaluation results on synthetic speech quality.}
\label{tab:evaluation_obj}
\scalebox{0.905}{
\begin{tabular}{l|ccc}
\hline 
Method                                & CER & WER & MOS \\ \hline \hline
\textit{Grund-truth}                      &  5.1~\%   &  17.9~\% & $4.16 \pm 0.11$   \\
\textit{Full-sentence}                    &  5.5~\%   & 18.2~\% & $3.82 \pm 0.11$  \\ \hline
\textit{Teacher}                          & 7.8~\%  & 22.2~\%  & $3.51 \pm 0.15$  \\
\textit{Unicontext}                       & 20.8~\%  & 49.4~\% & $3.10 \pm 0.16$   \\
\textit{Student} (medium, $\lambda=1$)    &  8.4~\%   & 22.2~\% & $3.47 \pm 0.15$    \\
\textit{Student} (medium, $\lambda=0.95$) &  12.7~\%   & 33.8~\% & $3.39 \pm 0.15$ \\
\textit{Student} (small, $\lambda=1$)     &   8.5~\%  & 23.9~\% & - \\
\textit{Student} (large, $\lambda=1$)     &  8.3~\%   & 23.3~\% & -    \\ \hline
\end{tabular}
}
\end{table}

\subsection{Subjective evaluation of synthetic speech quality}
As a subjective evaluation, we conducted a mean opinion score (MOS) evaluation test~\cite{steijl16mos} on naturalness.
We compared the methods described above except for ``\textit{Student} (small, $\lambda=1$)'' and ``\textit{Student} (large, $\lambda=1$)'', whose error rates were equivalent to those of ``\textit{Student} (medium, $\lambda=1$)''.
Forty listeners recruited through Amazon Mechanical Turk~\cite{mturk} participated in the evaluation, and each listener evaluated 30~speech samples, where we randomly chose five samples from the output utterances of the test data for each method.

Table~\ref{tab:evaluation_obj} shows the average MOS with 95~\% confidence intervals.
The MOS of ``\textit{Student} ($\lambda=1$)'' was significantly higher than that of \textit{Unicontext}, as were the results of the objective evaluation experiment.
The difference between the scores of \textit{Teacher} and \textit{Student} was 0.04, indicating that the proposed knowledge distillation method can make the quality of the lightweight student model very close to the teacher model.
However, since there was still a quality gap between \textit{Teacher} and \textit{Full-sentence}, further research is needed to improve the model structure of the teacher model.

\vspace{-1mm}
\section{Conclusions}
\vspace{-1mm}
We proposed a knowledge distillation method for efficiently estimating contextual embedding while using the linguistic knowledge of a large pre-trained language model.
The proposed training method using the teacher-student loss between the context embedding vectors distills knowledge from a GPT2-based teacher model into a lightweight student model.
\revise{Experimental results show that the proposed method achieves ten times faster inference than the conventional method without degrading synthetic speech quality and performs incremental synthesis much faster than the average speaking speed of human English speakers, demonstrating the availability of our method to real-time applications.}
Our future work will focus on the model architecture for incremental TTS to make its synthetic speech quality equivalent to that of sentence-level TTS methods.

\section{Acknowledgements}
Part of this work was supported by JSPS KAKENHI Grant Number 17H06101 (system development) and 18K18100 (system development), the MIC/SCOPE \#182103104 (core idea), and JST, Moonshot R\&D Grant Number JPMJPS2011 (experiment).

\bibliographystyle{IEEEbib}
\bibliography{tts}

\end{document}